\documentclass[12pt]{article}
\usepackage{latexsym}
\usepackage{tikz}
\usepackage{amsfonts}
\usepackage{amsmath}
\usepackage{tkz-graph}
\usepackage{amssymb}
\usepackage{comment}

\usepackage{hyperref}

\usepackage{tikz}
\newcommand*\circled[1]{\tikz[baseline=(char.base)]{
		\node[shape=circle,draw,inner sep=2pt] (char) {#1};}}
\usetikzlibrary{shapes}
\usetikzlibrary{snakes}
\usetikzlibrary{shapes}
\usetikzlibrary{arrows}
\usetikzlibrary{decorations.pathmorphing,positioning, shapes.geometric}

\newtheorem{Theorem}{Theorem}[section]

\newtheorem{Lemma}[Theorem]{Lemma}
\newtheorem{Observation}[Theorem]{Observation}
\newtheorem{Conjecture}{Conjecture}[section]
\newtheorem{Problem}{Problem}[section]

\newcommand{\join}{\; \circled{1} \;} 
\newcommand{\cojoin}{\; \circled{0} \;}

\newcommand{\qed}{\hfill $\Box$}

\def\inst#1{$^{#1}$}

\DeclareMathOperator{\cwd}{cwd}

\title {On the structure of {($4K_1$, $C_4$, $P_6$)}-free graphs}

\author{
	Ch\'inh T. Ho\`ang\inst{1}
	\and Ramin Javadi\inst{2}
	\and Nicolas Trotignon\inst{3} 
}
\date{}

\begin{document}
	\maketitle
	\begin{center}
		{\footnotesize
			
			\inst{1} Department of Computer Science and Physics, Wilfrid Laurier	University, \\Waterloo, Ontario, Canada
			
			\inst{2} Department of Mathematical Sciences, Isfahan University of Technology, 
			Iran \\
			\inst{3}  CNRS, ENS de Lyon, Université Lyon 1, LIP UMR 5668, 69342 Lyon Cedex 07, France.
		}
		
	\end{center}
	
	\begin{abstract}
		Determining the complexity of colouring ($4K_1, C_4$)-free graph is a long open problem.
		Recently Penev showed that there is a polynomial-time algorithm to colour a ($4K_1, C_4, C_6$)-free graph. In this paper, we will prove that if $G$ is a ($4K_1, C_4, P_6$)-free graph that contains a $C_6$, then $G$ has bounded clique-width. To this purpose, we use a new method to bound the clique-width, that is of independent interest. As a consequence, there is a polynomial-time algorithm to colour ($4K_1, C_4, P_6$)-free graphs.
	\end{abstract}

	\section{Introduction}
	
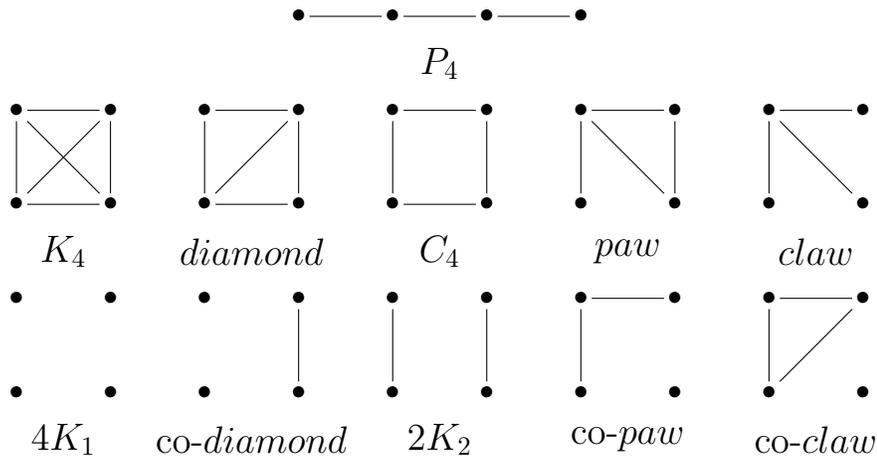
\begin{figure}
		\begin{center}
			\begin{tikzpicture} [scale = 1.25]
				\tikzstyle{every node}=[font=\small]
				
				\newcommand{\size}{1}
				
				\newcommand{\p4}{1}{
					\path (\size * 4, 0) coordinate (g1);
					\path (g1) +(-\size, 0) node (g1_1){};
					\path (g1) +(0, 0) node (g1_2){};
					\path (g1) +(\size, 0) node (g1_3){};
					\path (g1) +(2 * \size, 0) node (g1_4){};
					\foreach \Point in {(g1_1), (g1_2), (g1_3), (g1_4)}{
						\node at \Point{\textbullet};
					}
					\draw   (g1_1) -- (g1_2)
					(g1_2) -- (g1_3)
					(g1_3) -- (g1_4);
					\path (g1) ++(\size  / 2,-\size / 2) node[draw=none,fill=none] { {\large $P_4$}};
				} 
				
				\newcommand{\kfour}{2}{
					\path (0, - \size * 2) coordinate (g2);
					\path (g2) +(0, 0) node (g2_1){};
					\path (g2) +(0, \size) node (g2_2){};
					\path (g2) +(\size, \size) node (g2_3){};
					\path (g2) +(\size, 0) node (g2_4){};
					\foreach \Point in {(g2_1), (g2_2), (g2_3), (g2_4)}{
						\node at \Point{\textbullet};
					}
					\draw   (g2_1) -- (g2_2)
					(g2_1) -- (g2_3)
					(g2_1) -- (g2_4)
					(g2_2) -- (g2_3)
					(g2_2) -- (g2_4)
					(g2_3) -- (g2_4);
					\path (g2) ++(\size  / 2,-\size / 2) node[draw=none,fill=none] { {\large $K_4$}};
				}
				
				\newcommand{\diam}{3}{
					\path (\size * 2, - \size * 2) coordinate (g3);
					\path (g3) +(0, 0) node (g3_1){};
					\path (g3) +(0, \size) node (g3_2){};
					\path (g3) +(\size, \size) node (g3_3){};
					\path (g3) +(\size, 0) node (g3_4){};
					\foreach \Point in {(g3_1), (g3_2), (g3_3), (g3_4)}{
						\node at \Point{\textbullet};
					}
					\draw   (g3_1) -- (g3_2)
					(g3_1) -- (g3_3)
					(g3_1) -- (g3_4)
					(g3_2) -- (g3_3)
					(g3_3) -- (g3_4);
					\path (g3) ++(\size  / 2,-\size / 2) node[draw=none,fill=none] { {\large $diamond$}};
				}
				
				\newcommand{\cfour}{4}{
					\path (\size * 4, - \size * 2) coordinate (g4);
					\path (g4) +(0, 0) node (g4_1){};
					\path (g4) +(0, \size) node (g4_2){};
					\path (g4) +(\size, \size) node (g4_3){};
					\path (g4) +(\size, 0) node (g4_4){};
					\foreach \Point in {(g4_1), (g4_2), (g4_3), (g4_4)}{
						\node at \Point{\textbullet};
					}
					\draw   (g4_1) -- (g4_2)
					(g4_1) -- (g4_4)
					(g4_2) -- (g4_3)
					(g4_3) -- (g4_4);
					\path (g4) ++(\size  / 2,-\size / 2) node[draw=none,fill=none] { {\large $C_4$}};
				}
				
				\newcommand{\paw}{5}{
					\path (\size * 6, - \size * 2) coordinate (g5);
					\path (g5) +(0, 0) node (g5_1){};
					\path (g5) +(0, \size) node (g5_2){};
					\path (g5) +(\size, \size) node (g5_3){};
					\path (g5) +(\size, 0) node (g5_4){};
					\foreach \Point in {(g5_1), (g5_2), (g5_3), (g5_4)}{
						\node at \Point{\textbullet};
					}
					\draw   (g5_1) -- (g5_2)
					(g5_2) -- (g5_3)
					(g5_2) -- (g5_4)
					(g5_3) -- (g5_4);
					\path (g5) ++(\size  / 2,-\size / 2) node[draw=none,fill=none] { {\large $paw$}};
				}
				
				\newcommand{\claw}{6}{
					\path (\size * 8, - \size * 2) coordinate (g6);
					\path (g6) +(0, 0) node (g6_1){};
					\path (g6) +(0, \size) node (g6_2){};
					\path (g6) +(\size, \size) node (g6_3){};
					\path (g6) +(\size, 0) node (g6_4){};
					\foreach \Point in {(g6_1), (g6_2), (g6_3), (g6_4)}{
						\node at \Point{\textbullet};
					}
					\draw   (g6_1) -- (g6_2)
					(g6_2) -- (g6_4)
					(g6_2) -- (g6_3);
					\path (g6) ++(\size  / 2,-\size / 2) node[draw=none,fill=none] { {\large $claw$}};
				}
				
				\newcommand{\cokfour}{7}{
					\path (0, - \size * 4) coordinate (g7);
					\path (g7) +(0, 0) node (g7_1){};
					\path (g7) +(0, \size) node (g7_2){};
					\path (g7) +(\size, \size) node (g7_3){};
					\path (g7) +(\size, 0) node (g7_4){};
					\foreach \Point in {(g7_1), (g7_2), (g7_3), (g7_4)}{
						\node at \Point{\textbullet};
					}
					
					\path (g7) ++(\size  / 2,-\size / 2) node[draw=none,fill=none] { {\large $4K_1$}};
				}
				
				\newcommand{\codiamond}{8}{
					\path (\size * 2, - \size * 4) coordinate (g8);
					\path (g8) +(0, 0) node (g8_1){};
					\path (g8) +(0, \size) node (g8_2){};
					\path (g8) +(\size, \size) node (g8_3){};
					\path (g8) +(\size, 0) node (g8_4){};
					\foreach \Point in {(g8_1), (g8_2), (g8_3), (g8_4)}{
						\node at \Point{\textbullet};
					}
					\draw   (g8_3) -- (g8_4);
					\path (g8) ++(\size  / 2,-\size / 2) node[draw=none,fill=none] { {\large co-$diamond$}};
				}
				
				\newcommand{\cocfour}{9}{
					\path (\size * 4, - \size * 4) coordinate (g8);
					\path (g8) +(0, 0) node (g8_1){};
					\path (g8) +(0, \size) node (g8_2){};
					\path (g8) +(\size, \size) node (g8_3){};
					\path (g8) +(\size, 0) node (g8_4){};
					\foreach \Point in {(g8_1), (g8_2), (g8_3), (g8_4)}{
						\node at \Point{\textbullet};
					}
					\draw (g8_1) -- (g8_2)
					(g8_3) -- (g8_4);
					\path (g8) ++(\size  / 2,-\size / 2) node[draw=none,fill=none] { {\large $2K_2$}};
				}
				
				\newcommand{\copaw}{10}{
					\path (\size * 6, - \size * 4) coordinate (g9);
					\path (g9) +(0, 0) node (g9_1){};
					\path (g9) +(0, \size) node (g9_2){};
					\path (g9) +(\size, \size) node (g9_3){};
					\path (g9) +(\size, 0) node (g9_4){};
					\foreach \Point in {(g9_1), (g9_2), (g9_3), (g9_4)}{
						\node at \Point{\textbullet};
					}
					\draw   (g9_1) -- (g9_2)
					(g9_2) -- (g9_3);
					\path (g9) ++(\size  / 2,-\size / 2) node[draw=none,fill=none] { {\large co-$paw$}};
				}
				
				\newcommand{\coclaw}{11}{
					\path (\size * 8, - \size * 4) coordinate (g10);
					\path (g10) +(0, 0) node (g10_1){};
					\path (g10) +(0, \size) node (g10_2){};
					\path (g10) +(\size, \size) node (g10_3){};
					\path (g10) +(\size, 0) node (g10_4){};
					\foreach \Point in {(g10_1), (g10_2), (g10_3), (g10_4)}{
						\node at \Point{\textbullet};
					}
					\draw   (g10_1) -- (g10_2)
					(g10_1) -- (g10_3)
					(g10_2) -- (g10_3);
					\path (g10) ++(\size  / 2,-\size / 2) node[draw=none,fill=none] { {\large co-$claw$}};
				}
				
			\end{tikzpicture}
		\end{center}
		\caption{All four-vertex graphs}\label{fig:4vertex-graphs}
	\end{figure}

	Let $L$ be a list (class) of graphs. A graph $G$ is \emph{$L$-free} if $G$ does not contain an induced subgraph isomorphic to a graph in $L$. There has been much interest in finding polynomial-time algorithms to colour $L$-free graphs for several well-defined classes $L$. In this context, perhaps one of the most well-known open problems is to colour ``even hole-free'' graphs in polynomial time (definitions will be given later). Recently, many researchers studied graph classes  whose forbidden list $L$ contains graphs with four vertices. Figure~\ref{fig:4vertex-graphs} shows all 11 four-vertex graphs. Recent papers of Lozin and Malyshev \cite{Lozin}, and of Fraser, Hamel, Ho\`ang, Holmes and LaMantia \cite{FraHam2017}  discuss the state of the art on this problem,  identifying three outstanding classes: $L=(4K_1$, claw),
	$L=(4K_1$, claw, co-diamond), and $L=(4K_1, C_4$). For now, the solutions to these three problems seem elusive. A natural way to make progress is to enlarge the forbidden list $L$. Then the question is, which $L$ is an interesting list to be considered? In this paper, we study the class $L=\{4K_1, C_4\}$. Before introducing our problem, we will need to introduce a few definitions.
	
	%
	
	By $C_k$ (respectively, $P_k$), we denote the chordless cycle (respectively, path) on $k$ vertices. A {\it hole} is a $C_k$ with $k \geq 4$.  A hole is \emph{even} if it has an even  number of vertices.  The problem of colouring even-hole-free graphs has been extensively studied. 
	The combined results of Addario-Berry, Chudnovsky, Havet, Reed, and Seymour~\cite{addarioBerryEtAl:ehf} and Chudnovky and Seymour \cite{ChuSey2023}
	shows that for an even-hole-free graph $G$, the chromatic number of $G$ is at most two times its clique number minus one\footnote{[2] gave a proof of this result but it was later discovered that the proof is incomplete, [5] gave a correct proof that uses several results from [2]} (the clique number of a graph is the number of vertices in a largest clique). 
	It is currently not known whether even-hole-free graphs can be coloured in polynomial time. 
	However, there are polynomial-time algorithms for colouring some subclasses of even hole-free graphs  \cite{FraHam2018,CamCha2018,CamDas2018,maffray2019coloring}. See \cite{KloVus2009} for a survey on even hole-free graphs. Until now, we have discussed four colouring problems which are summarized as follows.

	\begin{Problem}\label{pro:claw}
		What is the complexity of colouring ($4K_1$, claw)-free graphs?
	\end{Problem}
	\begin{Problem}\label{pro:co-diamond}
		What is the complexity of colouring ($4K_1$, claw, co-diamond)-free graphs?
	\end{Problem}

	\begin{Problem}\label{pro:C4}
		What is the complexity of colouring ($4K_1, C_4$)-free graphs?
	\end{Problem}
	
	\begin{Problem}\label{pro:even-hole}
		What is the complexity of colouring even-hole-free graphs?
	\end{Problem}

	Problems \ref{pro:claw} and \ref{pro:co-diamond} have been studied by many researchers (see \cite{AbuHoa2022,DaiFol2022,DaiFol2022b, FraHam2018,DBLP:journals/algorithmica/PreissmannRT21}). This paper is concerned with Problem~\ref{pro:C4}.

	In \cite{FolFra2020}, Foley et al. proposed the intersection of Problems \ref{pro:C4} and \ref{pro:even-hole}. They asked, what is the complexity of colouring ($4K_1, C_4, C_6$)-free graphs. Penev (\cite{Pen2020,Pen2022}) answered this question by providing a polynomial time algorithm to colour ($4K_1, C_4, C_6$)-free graphs. 
	
	\begin{Theorem}[\cite{Pen2020}]\label{thm:penev}
		There is a polynomial-time algorithm to colour ($4K_1, C_4, C_6$)-free graphs. 
	\end{Theorem}
	
	Inspired by this result, we asked whether the $C_6$ in the list $L$ may be replaced by another graph. The $C_k$ is usually studied in conjunction with $P_k$. Thus, we chose $P_6$ (the chordless path on six vertices). That is, we will  prove the following theorem. 
	
	\begin{Theorem}\label{thm:main}
		There is a polynomial-time algorithm to color ($4K_1, C_4, P_6$)-free graphs.
	\end{Theorem}

	Our approach to prove Theorem~\ref{thm:main} is by bounding the clique-width of ($4K_1, C_4, P_6$)-free graphs that contain a~$C_6$.  We use a new method for bounding the clique-width that is of independent interest.  
	
	\subsection*{Outline of the paper}
	
	In Section \ref{sec:intro}, we will provide the necessary definitions and discuss the context of our problem (in particular we explain why to prove Theorem~\ref{thm:main} it is enough to bound the clique-width of ($4K_1, C_4, P_6$)-free graphs that contain a $C_6$ and explain our new upper bound on the clique-width). 
	In Section~\ref{sec:neigh}, we state and prove several observations about how a $C_6$ sees the rest of the graph in some ($4K_1, C_4, P_6$)-free graph. 
	In Section~\ref{sec:main}, we prove that a ($4K_1, C_4, P_6$)-free graph that contains a $C_6$ has bounded clique-width. 
	Finally, in Section~\ref{sec:conclusions}, we discuss further different methods to bound the clique-width and propose some related open problems. 
	
	\section{Background and method}\label{sec:intro}
	
	If a vertex $a$ is adjacent to a vertex $b$ in some graph $G$, then we write $a \sim b$. By $N(x)$, we denote the set vertices adjacent to $x$ and for any set $X$, $N_X(v)$ stands for the set $N(v) \cap X$. 
	The colouring problem is the problem of assigning a minimum number of colours to the vertices of an input graph so that two vertices receive different colours if they are adjacent. 
	It is well known that the colouring problem for a general graph is NP-hard. 
	
	Consider the following operations to build a labelled graph, starting from an empty graph.
	
	\begin{itemize}
		
		\item Create a vertex $u$ with label $i$. 
		
		\item  Take the disjoint union of two previously constructed labelled graphs.
		
		\item Join by an edge all vertices with label $i$ to all
		vertices with label $j$ for $i \not= j$.
		
		\item  Relabel all vertices with label $i$ by label $j$.
	\end{itemize}
	
	The {\it clique-width} of an unlabelled graph $G$, is the minimum number of labels needed to build the graph with the above four operations.  
	For instance, the reader may verify that for any $n\geq 2$, the clique on $n$ vertices has clique-width 2, and for any $n\geq 3$,  $P_n$ has clique-width~3. Rao \cite{Rao} proved the following theorem. 
	
	\begin{Theorem}[\cite{Rao}]\label{thm:Rao2007}
		The colouring problem is polynomial-time solvable for graphs with bounded clique-width.
	\end{Theorem}
	
	A vertex in a graph is {\it simplicial} if its neighbourhood is a clique. If $x$ is a simplicial vertex, then $\chi(G) = \max (\chi(G-x), |N(x)| + 1)$; thus if we know $\chi(G-x)$, then we know $\chi(G)$. Penev proved the following.
	
	\begin{Theorem}[\cite{Pen2020}]\label{thm:n-simplicial}
		The class of ($4K_1, C_4, C_6, C_7$)-free graphs that do not contain a simplicial vertex has bounded clique-width. 
	\end{Theorem}
	
	Foley, Fraser, Ho\`ang, Holmes, and LaMantia proved the following. 
	
	\begin{Theorem}[\cite{FolFra2020}]\label{thm:C7}
		The class of ($4K_1, C_4, C_6$)-free graphs that contains a $C_7$ has bounded clique-width. 
	\end{Theorem}
	
	Theorems~\ref{thm:C7}, \ref{thm:n-simplicial} and~\ref{thm:Rao2007} imply that the colouring problem for ($4K_1, C_4, C_6$)-free graphs can be solved in polynomial time as already stated in Theorem~\ref{thm:penev}. We will prove:
	
	\begin{Theorem}\label{thm:main-bounded}
		The class of ($4K_1, C_4, P_6$)-free graphs that contains a $C_6$ has bounded clique-width (at most~27).
	\end{Theorem}
	
	It is easily seen that Theorems~\ref{thm:penev}, \ref{thm:main-bounded} and~\ref{thm:Rao2007} imply Theorem~\ref{thm:main}.  In Theorem~\ref{thm:main-bounded}, it is essential to assume that $G$ contains a $C_6$ because there exist ($4K_1, C_4, P_6$)-free graphs with arbitrarily large clique-width (we postpone their description to the end of the section because we need some terminology to describe them). 
	
	Let us now explain our tool to bound the clique-width. 
	
	\subsection*{Bounding the clique-width}
	
	Given disjoint  sets $X,Y$ of vertices in some graph $G$, we say that $X$ is {\it complete} (resp. {\it anticomplete}) to $Y$ if there are all possible edges (resp. no edge) between vertices in $X$ and vertices in $Y$; we denote this relationship by $X\join Y$ (resp. $X\cojoin Y$).  
	Also, we say that $X$ is {\it uniform} to $Y$ if $X \join Y$ or $X \cojoin Y$.   
	We say that $X$ is {\it monotone} to $Y$ if for all $u, v\in X$, either $N_Y(u) \subseteq N_Y(v)$ or $N_Y(v) \subseteq N_Y(u)$. Observe that the empty set is complete, anticomplete, uniform and monotone to any set of vertices of any graph. 
	The following well-known observation is easy and we omit its proof. 
	
	\begin{Theorem}\label{th:monotone}
		For any graph $G$ and every disjoint sets $X, Y \subset V(G)$, the following are equivalent.
		\begin{itemize}
			\item $X$ is monotone to $Y$.
			\item $Y$ is monotone to $X$.
			\item There does not exist $x, x' \in X$ and $y, y'\in Y$ such that $x\sim y, x'\sim y'$, $x\not\sim y'$ and $x' \not\sim y$. 
			\item The vertices of $X$ (resp.\ $Y$) are linearly ordered by the inclusion of their neighbourhood in $Y$ (resp. $X$). 
		\end{itemize}
		
		In particular, if $X$ and $Y$ are two disjoint cliques in a $C_4$-free graph, then $X$ is  monotone to $Y$. 
	\end{Theorem}

	Let $G$ be a graph and ${\cal P} = (A_1, \ldots, A_k$) be a partition of $V(G)$. 
	We say that $\mathcal P$ is \emph{monotone} if for all $i\in \{1, \dots, k\}$, $A_i$ is monotone to $V(G) \setminus A_i$.  
	In a monotone partition, every set $A_i$ is linearly ordered with respect to the inclusion of neighbourhoods outside of $A_i$ (see Theorem~\ref{th:monotone}). So, in each non-empty set $A_i$, there exists at least one \emph{maximal} (resp.\ \emph{minimal}) vertex, that is a vertex $v$ such that $N_{V(G) \setminus A_i}(v)$ is inclusion-wise maximal (resp.\ minimal). 
	A vertex in $A_i$ that is either maximal or minimal is \emph{extreme}. 
	We say that a monotone partition $\mathcal P$ has the \emph{extreme vertex property} if there exists $i\in \{1, \dots, k\}$ such that some extreme vertex $v$ of $A_i$ is uniform to all sets $A_j$ for all $j\in \{1, \dots, k\} \setminus \{i\}$. 
	We say that $\mathcal P$ has the \emph{hereditary extreme vertex property} (\emph{hev-property} for short) if for every set $X\subseteq V(G)$, the partition $\mathcal P[X]$ has the extreme vertex property in $G[X]$ (where $\mathcal P[X] = (A_1\cap X, \dots, A_k \cap X)$ and $G[X]$ is the subgraph of $G$ induced by $X$).

	\begin{Theorem} 
		\label{th:bcw}
		Let $G$ be a graph and $\mathcal P = (A_1, \dots, A_k)$ be a monotone partition of $V(G)$ into cliques $A_1, \ldots, A_k$.  If $\mathcal P$ has the hev-property, then $G$ has clique-width at most $k+1$. 
	\end{Theorem}
	
	\noindent {\it Proof.} 
	Let us prove by induction on $|V(G)|$ that $G$ can be constructed with $k+1$ labels in such a way that at the end of the construction, for all $j\in \{1, \dots, k\}$,  every vertex in $A_j$ has label $L_{j}$.  
	This is clear if $|V(G)| = 1$, so we suppose that $|V(G)| \geq 2$.  
	From the definition of the hev-property, consider an extreme vertex $v$ in some set~$A_i$, that is uniform to all $A_j$ with $j\neq i$. 
	By the heredity property, $\mathcal P[V(G) \setminus \{v\}]$ has the hev-property for $G-v$. So, by the induction hypothesis $G-v$ can be constructed as stated above. 
	Now give the label $L$ to $v$ ($L$ is a special label used only for $v$).  
	
	For every set $A_{j}\setminus \{v\}$ ($j\in \{1, \dots, k\}$) that is complete to $v$,  
	make the label~$L$ complete to the label $L_{j}$.  Then relabel $v$ with the label $L_{i}$. This completes the construction.
	\qed
	
	The following is a convenient way to prove the hev-property. 
	\begin{Lemma}
		\label{l:deg1B}
		Let $G$ be a graph and $\mathcal P = (A_1, \dots, A_k)$ be a monotone partition of $V(G)$ such that $A_1$ is uniform to all sets $A_j$ ($j\in \{2,\dots, k\})$ except possibly one. If $(A_2, \dots, A_k)$ has the hev-property (for $G[A_2 \cup \dots \cup A_k]$), then $\mathcal P$ has the hev-property (for $G$). 
	\end{Lemma}
	
	\noindent {\it Proof.} 
	If $k=1$, then the conclusion is clear, so suppose $k\geq 2$. 
	Let us prove that some extreme vertex of some set $A_i$ ($i\in \{1, \dots, k\}$) is uniform to all sets $A_j$, $j\neq i$. 
	
	From the assumption, there exists some set $A_i$ (with $i\in \{2, \dots, k)$) such that some extreme vertex $v$ in $A_i$ is uniform to all sets $A_j$ ($j\geq 2$, $j\neq i$).  
	If $v$ is uniform to $A_1$ in $G$, we are done, so we may assume it is not.  
	So, $v$ has neighbours and non-neighbours in $A_1$. By the assumption, $A_1$ is uniform to all sets $A_j$, $j\geq 2, j\neq i$. 
	If $v$ is a maximal vertex of $A_i$, then a minimal vertex of $A_1$ must be
	anticomplete to $A_i$, because otherwise, by the maximality of $v$, $v$ should be complete to $A_1$, a contradiction.  
	If $v$ is a minimal vertex of $A_i$, then a symmetric argument shows that a maximal vertex of $A_1$ is complete to $A_i$.  
	So we have found in $A_1$ an extreme vertex $w$ that is uniform to all sets $A_i$ ($i\in \{2, \dots, k\}$).  
	
	The same proof can be done in $G[X]$ with the partition $\mathcal P[X]$ for any set $X\subseteq V(G)$, so $G$ has the hev-property. 
	\qed
	
	\medskip
	
	The following lemma will be useful for us later. 
	
	\begin{Lemma}\label{l:uniMon}
		Let $G$ be a graph and ${\cal P} = (A_1, \ldots, A_k$) be a partition of $V(G)$. If $A_1$ is  monotone to all sets $A_j$ ($j\in \{2, \dots, k\}$), and uniform to all sets $A_j$ ($j\in \{2, \dots, k\}$) except possibly one, then $A_1$ is monotone to $V(G) \setminus A_1$. 
	\end{Lemma}
	
	\noindent{\it Proof.}
	Otherwise, by Theorem~\ref{th:monotone}, there exist $x, x' \in A_1$ and $y, y'\notin A_1$ such that $x\sim y, x'\sim y'$, $x\not\sim y'$ and $x' \not\sim y$.  Since $A_1$ is uniform to all sets $A_j$ where $j\in \{2, \dots, k\}$ except possibly one, the two vertices $y$ and $y'$ must be in the same set $A_j$.  Since by assumption $A_1$ is monotone to $A_j$, there is a contradiction. 
	\qed
	
	\subsection*{Construction of $(4K_1, C_4, P_6)$-free graphs of large clique-width}
	
	As announced above, let us  describe $(4K_1, C_4, P_6)$-free graphs of arbitrarily large clique-width. We rely on a construction of the first and last author of the present paper, see~\cite{{DBLP:journals/dm/HoangT24}}.  
	They construct graphs of unbounded clique-width that admit a monotone partition into three cliques, called \emph{3-rings}. This is enough, because it is easy to check that none of $4K_1$, $C_4$ and $P_6$ admits a monotone partition into three cliques. Hence, the 3-rings  are $(4K_1, C_4, P_6)$-free.

	\section{Neighbourhood of the $C_6$}
	\label{sec:neigh}
	
	In this section, $G$ is a ($4K_1, C_4, P_6$)-free graph.
	We assume that $G$ contains a $C_6$ as an induced subgraph and examine its neighbourhood.  
	Enumerate the vertices of the $C_6$ as $c_1, c_2, \ldots , c_6$ such that $c_i \sim c_{i+1}$  with the subscripts taken modulo $6$ (for the rest of the paper, when the context is clear, the subscripts will be understood to be taken modulo 6). 
	For $i=0, \dots, 6$, let $X_i$ be the set of vertices in $V(G)\setminus\{c_1,\dots, c_6\}$ that have exactly $i$ neighbours in $\{c_1, \dots, c_6\}$. Obviously, $V(G) = \{c_1, \dots, c_6\} \cup \bigcup_{i=0}^{i=6} X_i$.

	\begin{Observation}\label{obs:X-empty}
		We have $X_0 \cup X_1 \cup X_5 = \emptyset$. 
	\end{Observation}
	\noindent{\it Proof.} If $X_0 \not= \emptyset$ or $X_1 \not= \emptyset$, then $G$ contains an induced $4K_1$. If $X_5 \not= \emptyset$, then $G$ contains an induced $C_4$. \qed
	
	Now, we know that $V(G) = \{c_1, \dots, c_6\} \cup  X_2 \cup X_3 \cup X_4 \cup X_6$.  Let $X_{2, j}$ be the set  of vertices  $x \in X_2$ such that $x$ is adjacent to  $c_j$ and $c_{j+3}$.  Let $X_{3,j}$ be the set of vertices $x \in X_3$ such that $x$ is adjacent to  $c_j$, $c_{j+1}$ and  $c_{j+2}$.  Let $X_{4,j}$ be the set of vertices $x \in X_4$ such that $x$ is adjacent to  $c_j$, $c_{j+1}$, $c_{j+2}$ and $c_{j+3}$. 
	
	\begin{Observation}\label{obs:X2}
		If $x \in X_2$ then $x \in X_{2,j}$ for some $j$. Furthermore, $X_{2,j}$ is a clique. 
	\end{Observation}
	{\it Proof.} Otherwise, $G$ contains a $C_4$ or a $P_6$. \qed
	
	\begin{Observation}\label{obs:X3}
		If $x \in X_3$ then $x \in X_{3,j}$ for some $j$. Furthermore, $X_{3,j}$ is a clique.
	\end{Observation}
	{\it Proof.} Otherwise, $G$ contains a $C_4$. \qed
	
	\begin{Observation}\label{obs:X4}
		If $x \in X_4$ then $x \in X_{4,j}$ for some $j$. Furthermore, $X_{4,j}$ is a clique.
	\end{Observation}
	{\it Proof.} Otherwise, $G$ contains a $C_4$. \qed

	\subsection{Adjacency relationships of the set $X_6$}\label{sec:X6}
	
	\begin{Observation}\label{obs:AdjX6}
		$X_6$ is a clique and $X_{6} \join (V(G) \setminus X_6)$  (in particular, $X_6$ is monotone to $V(G) \setminus X_6$). 
	\end{Observation} 
	{\it Proof}. Otherwise, $G$ contains a $C_4$. \qed

	\subsection{Adjacency relationships of the set $X_2$}\label{sec:X2}
	\begin{Observation}\label{obs:X2-uniform} 
		For some integer $j$, $X_2 = X_{2, j}$ (in particular, $X_2$ is a clique).  Furthermore, for any set $Y \in \{X_{3,1}, \ldots, X_{3,6}, Y_{4,1}, \ldots , Y_{4,6}, X_6 \}$, $X_2$ is uniform to $Y$. In particular, $X_2$ is monotone to $V(G)\setminus X_2$ (in fact, all vertices in $X_2$ have the same neighborhood in $V(G)\setminus X_2$).
	\end{Observation}
	{\it Proof.} 
	Suppose that $u\in X_{2, j}$ and $v\in X_{2, j'}$ with $j\neq j'$.  Then $G$ contains a $C_4$ (if $u\sim v)$ or a $4K_1$ (otherwise).  So, for some $j$, $X_2=X_{2, j}$.  By Observation~\ref{obs:X2}, it follows that $X_2$ is a clique. 
	
	Up to symmetry, we assume that $X_2 = X_{2, 1}$.  It follows that $X_2$ is complete to $X_{3, 3} \cup X_{3, 6}$ (otherwise $G$ contains a $4K_1$), complete to $X_{4, 1} \cup X_{4, 4}$ (otherwise $G$ contains a $C_4$) and anticomplete to $X_{3, 1} \cup X_{3, 2} \cup X_{3, 4} \cup X_{3, 5} \cup X_{4, 2} \cup X_{4, 3} \cup X_{4, 5} \cup X_{4, 6}$ (otherwise $G$ contains a $C_4$). Also, by Observation~\ref{obs:AdjX6}, $X_2$ is complete to $X_6$. Therefore, for each vertex $u\in X_2$, the neighborhood of $u$ in $V(G)\setminus X_2$ is $X_{3,3}\cup X_{3,6}\cup X_{4, 1} \cup X_{4, 4} \cup X_6\cup \{c_1,c_2\}$.
	Hence,  $X_2$ is monotone to $V(G) \setminus X_2$. 
	\qed

	\subsection{Adjacency relationships of the set $X_4$}\label{sec:X4}
	
	\begin{Observation}\label{obs:X4i-X4j}
		For all $j\in \{1, \dots, 6\}$, $X_{4,j} \join (X_{4,j+1} \cup X_{4,j+3} \cup X_{4,j+5})$ and 
		$X_{4,j} \cojoin (X_{4,j+2} \cup X_{4,j+4})$. 
	\end{Observation}
	\noindent
	{\it Proof.}  Consider a vertex $x \in X_{4,j}$. 
	
	Let $y$ be a vertex in $X_{4,j+1} \cup X_{4,j+3} \cup X_{4,j+5}$. It is a routine matter to check that $x$ and $y$ have two common neighbours that are not adjacent. It follows that $x\sim y$ for otherwise, $G$ contains a $C_4$.  Hence,  $X_{4,j} \join (X_{4,j+1} \cup X_{4,j+3} \cup X_{4,j+5})$. 
	
	Let $y$ be a vertex in $X_{4,j+2} \cup X_{4,j+4}$. It is a routine matter to check that $x$ and $y$ are the ends of some $P_4$ with two internal vertices in the $C_6$. If follows that $x\not\sim y$,  for otherwise, $G$ contains a $C_4$.  Hence, $X_{4,j} \cojoin (X_{4,j+2} \cup X_{4,j+4})$. \qed
	
	\begin{Observation}\label{obs:X4-join-X3}
		For all $j\in \{1, \dots, 6\}$, we have $X_{4,j} \join (X_{3,j} \cup X_{3,j+1})$ and $X_{4,j} \cojoin (X_{3,j+3} \cup X_{3,j+4})$.
	\end{Observation}
	{\it Proof.} Let $x$ be a vertex in $X_{4,j}$ and $y$ be a vertex in  $X_{3,j}$. If $x \not\sim y$, then $\{x,c_j,y,c_{j+2}\}$ induces a $C_4$. So, $X_{4,j} \join X_{3,j}$. A symmetrical argument shows that $X_{4,j} \join  X_{3,j+1}$. 
	
	%
	%
	Now, let $x$ be a vertex in $X_{4,j}$ and $y$ be a vertex in  $X_{3,j+3}$. If $x \sim y$, then $\{x,c_j, c_{j+5}, y\}$ induces a $C_4$. So $X_{4,j} \cojoin  X_{3,j+3}$. A symmetrical argument shows that $X_{4,j} \cojoin  X_{3,j+4}$. \qed 
	
	\medskip 
	
	Note that $X_{4, j}$ may fail to be uniform to $X_{3, j+2}$ and $X_{3, j+5}$.

	\begin{Observation}\label{obs:X4partition}
		For all $j \in \{1, \dots, 6\}$, there exists a partition $(X_{4,j}^0, X_{4,j}^1)$ of $X_{4,j}$ such that $X_{4,j}^0 \join X_{3,j+5}$  and $X_{4,j}^1 \join X_{3,j+2}$.
	\end{Observation}
	{\it Proof.} Otherwise, there exists $x \in X_{4, j}$, $y\in X_{3, j+5}$ and $z\in X_{3, j+2}$ such that $x \not\sim y$ and $x\not\sim z$. So $\{y, c_i, x, c_{i+2}, z, c_{i+4}\}$ induces a $P_6$, a contradiction.\qed 
	
	\medskip
	
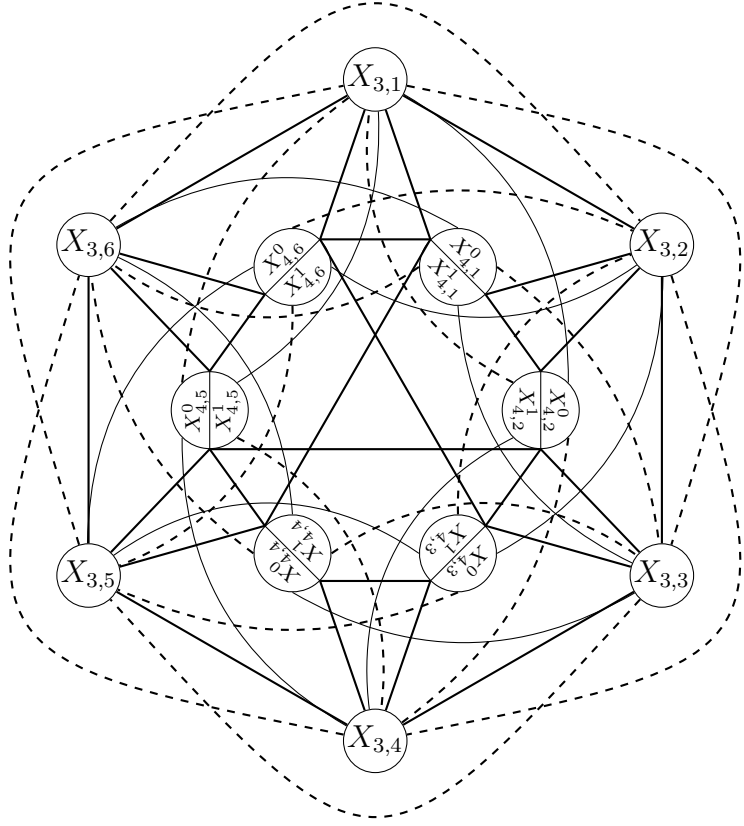
\begin{figure}[t]
		\vskip -4mm
		\centering
		\begin{tikzpicture}[scale=2.2,
			every node/.style={minimum size=8mm, inner sep=0pt},
			X3node/.style={circle, draw, minimum size=8mm, inner sep=0pt},
			X4node/.style={circle, minimum size=8mm, inner sep=0pt}
			]
			
			\foreach \i in {1,...,6}{
				\node[X3node] (X3\i) at (150-60*\i:2) {$X_{3,\i}$};
			}
			
			\node[X3node,rotate=45] (X46) at (120-60*0:1) {$\stackrel{X^0_{4,6}}{_{X^1_{4,6}}}$};
			\node[X3node,rotate=-45] (X41) at (120-60*1:1) {$\stackrel{X^0_{4,1}}{_{X^1_{4,1}}}$};
			\node[X3node,rotate=-90] (X42) at (120-60*2:1) {$\stackrel{X^0_{4,2}}{_{X^1_{4,2}}}$};
			\node[X3node,rotate=225] (X43) at (120-60*3:1) {$\stackrel{X^0_{4,3}}{_{X^1_{4,3}}}$};
			\node[X3node,rotate=135] (X44) at (120-60*4:1) {$\stackrel{X^0_{4,4}}{_{X^1_{4,4}}}$};
			\node[X3node,rotate=90] (X45) at (120-60*5:1) {$\stackrel{X^0_{4,5}}{_{X^1_{4,5}}}$};
			
		\draw[thick] (X31) -- (X32);
		\draw[thick] (X32) -- (X33);
		\draw[thick] (X33) -- (X34);
		\draw[thick] (X34) -- (X35);
		\draw[thick] (X35) -- (X36);
		\draw[thick] (X36) -- (X31);
		\draw (X41.west) -- (X41.east);
		\draw (X42.west) -- (X42.east);
		\draw (X43.west) -- (X43.east);
		\draw (X44.east) -- (X44.west);
		\draw (X45.west) -- (X45.east);
		\draw (X46.west) -- (X46.east);
		\draw[thick] (X41.east) -- (X42.west);
		\draw[thick] (X42.east) -- (X43.west);
		\draw[thick] (X43.east) -- (X44.west);
		\draw[thick] (X44.east) -- (X45.west);
		\draw[thick] (X45.east) -- (X46.west);
		\draw[thick] (X46.east) -- (X41.west);
		\draw[thick] (X31) -- (X41.west);
		\draw[thick] (X31) -- (X46.east);
		\draw[thick] (X32) -- (X41.east);
		\draw[thick] (X32) -- (X42.west);
		\draw[thick] (X33) -- (X42.east);
		\draw[thick] (X33) -- (X43.west);
		\draw[thick] (X34) -- (X43.east);
		\draw[thick] (X34) -- (X44.west);
		\draw[thick] (X35) -- (X44.east);
		\draw[thick] (X35) -- (X45.west);
		\draw[thick] (X36) -- (X45.east);
		\draw[thick] (X36) -- (X46.west);
		\draw[thick] (X41.west) -- (X44.east);
		\draw[thick] (X42.east) -- (X45.west);
		\draw[thick] (X43.west) -- (X46.east);
		\draw (X41.north west) to[bend right=30] (X36);
		\draw (X41.south east) to[bend right=30] (X33);
		\draw (X42.north west) to[bend right=30] (X31);
		\draw (X42.south east) to[bend right=35] (X34);
		\draw (X43.north west) to[bend right=30] (X32);
		\draw (X43.south east) to[bend right=35] (X35);
		\draw (X44.north west) to[bend right=30] (X33);
		\draw (X44.south east) to[bend right=30] (X36);
		\draw (X45.north west) to[bend right=30] (X34);
		\draw (X45.south east) to[bend right=30] (X31);
		\draw (X46.north west) to[bend right=30] (X35);
		\draw (X46.south east) to[bend right=35] (X32);
		\draw[thick,dashed] (X41.north east) to[bend left=23] (X33);
		\draw[thick,dashed] (X42.north east) to[bend left=23] (X34);
		\draw[thick,dashed] (X43.north east) to[bend left=23] (X35);
		\draw[thick,dashed] (X44.north east) to[bend left=23] (X36);
		\draw[thick,dashed] (X45.north east) to[bend left=23] (X31);
		\draw[thick,dashed] (X46.north east) to[bend left=23] (X32);
		\draw[thick,dashed] (X41.south west) to[bend left=35] (X36);
		\draw[thick,dashed] (X42.south west) to[bend left=35] (X31);
		\draw[thick,dashed] (X43.south west) to[bend left=35] (X32);
		\draw[thick,dashed] (X44.south west) to[bend left=35] (X33);
		\draw[thick,dashed] (X45.south west) to[bend left=35] (X34);
		\draw[thick,dashed] (X46.south west) to[bend left=35] (X35);
		\draw[thick,dashed] (X31) .. controls (2.5,1.5)  .. (X33);
		\draw[thick,dashed] (X33) .. controls (0,-2.9)  .. (X35);
		\draw[thick,dashed] (X35) .. controls (-2.5,1.5)  .. (X31);
		\draw[thick,dashed] (X32) .. controls (2.5,-1.5)  .. (X34);
		\draw[thick,dashed] (X34) .. controls (-2.5,-1.5)  .. (X36);
		\draw[thick,dashed] (X36) .. controls (0,2.9)  .. (X32);
		
		%
	\end{tikzpicture}
	\vskip -5mm
	\caption{Adjacencies in $X_3\cup X_4$. Solid line (resp. dashed line) between two sets $A$ and $B$ means that $A$ is complete (resp. monotone) to $B$.}
	\label{fig:X34}
\end{figure}

	From now on, we consider the following partition of $V(G)$:
	$$
	\mathcal P = \left(\{c_1\}, \dots, \{c_6\}, X_6, X_2, X_{4, 1}^0, X_{4, 1}^1, \dots, X_{4, 6}^0, X_{4, 6}^1, X_{3, 1}, \dots, X_{3, 6}\right).
	$$
	The adjacencies between sets in $\mathcal{P}$ are illustrated in Figure~\ref{fig:X34}. Some of these facts are already proved and some will be proved in the sequel. 
	
	\begin{Observation}\label{obs:X4Mon}
		For all $j \in \{1, \dots, 6\}$ and $\eta\in \{0, 1\}$, $X_{4,j}^\eta$ is uniform to all sets of $\mathcal P\setminus X_{4,j}^\eta$  except possibly one, and is monotone to $V(G) \setminus X_{4,j}^\eta$.
	\end{Observation}
	{\it Proof.}  
	The set $X_{4,j}^\eta$ is uniform to every set $X_{3, i}$ except possibly one because of Observations~\ref{obs:X4-join-X3} and~\ref{obs:X4partition}. It is uniform to all other sets of $\mathcal P$ by Observations~\ref{obs:X4}, \ref{obs:AdjX6}, \ref{obs:X2-uniform} and \ref{obs:X4i-X4j}. 
	
	By Lemma~\ref{l:uniMon}, to prove the second assertion, it is enough to prove that $X_{4,j}^\eta$  is monotone to all sets  $X_{3,i}$. This follows from Theorem~\ref{th:monotone}, since $X_{4,j}^\eta$ and $X_{3,i}$ are cliques in a $C_4$-free graph. 
	\qed 
	
\subsection{Adjacency relationships of the set $X_3$}\label{sec:X3}

\begin{Observation}\label{obs:X3-join-X3}
	For all $j\in \{1, \dots, 6\}$, we have $X_{3,j} \join (X_{3,j+1} \cup X_{3,j+5})$ and $X_{3,j} \cojoin X_{3,j+3} $.
\end{Observation}
\noindent
{\it Proof.} Let $x$ be a vertex in $X_{3,j}$ and $y$ be a vertex in  $X_{3,j+1}$. If $x \not\sim y$  then $\{x,c_{j+1}, y, c_{j+3}, c_{j+4},  c_{j+5}\}$ induces a $P_6$.  So, $X_{3,j} \join  X_{3,j+1}$. A symmetrical argument shows that $X_{3,j} \join  X_{3,j+5}$. 


Now, let $x$ be a vertex in $X_{3,j}$ and $y$ be a vertex in  $X_{3,j+3}$. If $x \sim y$, then $\{x,c_{j+2}, c_{j+3}, y \}$ induces a $C_4$. So, $X_{3,j} \cojoin X_{3,j+3} $. \qed

\medskip

Note that $X_{3, j}$ may fail to be uniform to $X_{3, j+2}$ and $X_{3, j+4}$.

\begin{Observation}\label{obs:X3P4}
	For all $j\in \{1, \dots, 6\}$, $X_{3, j}$ is monotone to $X_{3, j+2} \cup X_{3, j+4}$. 
\end{Observation}

\noindent

{\it Proof.}
Otherwise, there exists $x_j, x'_j\in X_{3, j}$,  $y, z \in X_{3, j+2} \cup X_{3, j+4}$, with $x_j\sim y$,  $x'_j \sim z$, $x_j\not\sim z$ and $x'_j \not\sim y$. If $y \sim z$, then $\{x_j, x'_j, y, z\}$ induces a $C_4$, a contradiction. So we may assume $y \not\sim z$. Since each of $X_{3, j+2},  X_{3, j+4}$ is a clique, we may assume $y \in X_{3,j+2}, z \in X_{3,j+4}$. Now, $\{c_{j+3}, y, x_j, x'_j, z, c_{j+5}\}$ induces a $P_6$, a contradiction.


\qed

\begin{Observation}\label{obs:X3Mon}
	For all $j\in \{1, \dots, 6\}$, $X_{3, j}$ is monotone to $V(G) \setminus X_{3, j}$. 
\end{Observation}

\noindent
{\it Proof.} 
Otherwise, there exists $x, x'\in X_{3, j}$ and $y, z\in V(G) \setminus X_{3, j}$ such that $x\sim y$,  $x' \sim z$, $x\not\sim z$ and $x' \not\sim y$. We have $y\not\sim z$ for otherwise
$\{x, x', y, z\}$ induces a $C_4$.	Note that $y$ (resp.\ $z$) cannot be in a set that is uniform to $X_{3, j}$. So let us first list the sets of $\mathcal P$ that are uniform to $X_{3, j}$. 

By definition of $X_{3, j}$, each set $\{c_i\}$ is uniform to $X_{3, j}$.  By Observations~\ref{obs:AdjX6} and~\ref{obs:X2-uniform}, $X_6$ and $X_2$ are uniform to  $X_{3, j}$. 
By Observations~\ref{obs:X4-join-X3} and~\ref{obs:X4partition},  $X_{3, j}$ is uniform to all sets $X_{4, i}^\eta$, except possibly $X_{4, j+1}^1$ and $X_{4, j+4}^0$.  By Observation~\ref{obs:X3-join-X3}, $X_{3, j}$ is uniform to all sets $X_{3, i}$ except possibly $X_{3, j+2}$ and $X_{3, j+4}$. 

To sum up, $y$ (resp. $z$) must be in one of the following sets: $X_{4, j+1}^1$, $X_{4, j+4}^0$,  $X_{3, j+2}$ or $X_{3, j+4}$. Note that since $y\not\sim z$ and all these sets are cliques, $y$ and $z$ must be in different sets.  By Observation~\ref{obs:X4i-X4j}, $X_{4, j+1}^1 \join X_{4, j+4}^0$. By Observation~\ref{obs:X4-join-X3}, $X_{4, j+1}^1 \join X_{3, j+2}$ and $X_{4, j+4}^0 \join X_{3, j+4}$.  So, up to a swap of $y$ and $z$, there are only three possibilities: 
\begin{itemize}
	\item $y \in X_{3,j+2}$, $z \in X_{3,j+4}$,  or  
	\item $y \in X_{4,j+1}$, $z \in X_{3,j+4}$, or
	\item $y \in X_{4,j+4}$, $z \in X_{3,j+2}$.    
\end{itemize}

The first case is impossible by Observation~\ref{obs:X3P4}. In the second case, $\{c_{j+3}, y, x, x', z, c_{j+5}\}$ induces a~$P_6$.  In the third case, $\{c_{j+5}, y, x, x', z, c_{j+3}\}$ induces a~$P_6$.  
\qed

\begin{Observation}\label{obs:triangular}
	For all $j\in \{1, \dots, 6\}$,  one of the following holds: 
	\begin{enumerate}
		\item For some $t$ in $\{j, j+2, j+4\}$, we have $X_{3, t} \cojoin (X_{3, t+2} \cup X_{3, t+4})$.
		\item There exists partitions	$(X_{3,j}^0, X_{3,j}^1)$ of $X_{3,j}$, $(X_{3,j+2}^0, X_{3,j+2}^1)$ of $X_{3,j+2}$ and $(X_{3,j+4}^0, X_{3,j+4}^1)$ of $X_{3,j+4}$ such that the three sets $X^1_{3,t}$ are all non-empty, and  for all distinct $t,r$ in $\{j, j+2, j+4\}$, we have  $X_{3,t}^1 \join X_{3,r}^1$  and $X_{3,t}^0 \cojoin X_{3,r}$.	    
	\end{enumerate}
\end{Observation}

\noindent
{\it Proof.} We  first establish the following fact. 
\begin{equation}\label{eq:triangle}
	\text{\parbox{.85\textwidth}{Let $\{ t,r,s\} = \{j,j+2, j+4\}$. If there are vertices $a \in X_{3,t} , \; b \in X_{3,r}, c \in X_{3,s}$ with $a \sim b, a \sim c$, then $b \sim c$.}}
\end{equation}
Suppose that $b \not\sim c$. For simplicity, we may assume $t=j, r=j+2, s = j+4$. Then $\{a,b,c_{j+4},c\}$ induces a $C_4$. So~\eqref{eq:triangle} holds. 

Let $j\in \{1, \dots, 6\}$.  Suppose that the first conclusion of the observation does not hold. Up to symmetry, it means that there exists an edge $x_j x_{j+2}$ with $x_j\in X_{3, j}$, $x_{j+2}\in X_{3, j+2}$ and an edge $x'_j x_{j+4}$  with $x'_j\in X_{3, j}$, $x_{j+4}\in X_{3, j+4}$. 
By Observation~\ref{obs:X3P4},  we may assume up to symmetry that $x_j\sim x_{j+4}$. By~\eqref{eq:triangle}, $\{x_j, x_{j+2}, x_{j+4}\}$ induces a triangle. 

For all $i \in \{j, j+2, j+4\}$, we set $X_{3,i}^{1} = N_{X_{3, i}}(x_{i+2})$ and $X_{3,i}^{0} =   X_{3,i}  \setminus X_{3,i}^{1}$.  Let us prove that these sets behave as claimed in the second conclusion of the observation. 

Let us prove that $X_{3,j}^1 \join X_{3,j+2}^1$. So, consider $x'_j \in X_{3,j}^1$ and $x'_{j+2} \in X_{3,j+2}^1$.  By definition, $x'_j \sim x_{j+2}$ and $x'_{j+2}\sim x_{j+4}$.  By \eqref{eq:triangle} applied to $x_{j+2}$, $x'_j$ and $x_{j+4}$, we have  $x_{j+4} \sim x'_j$.  By \eqref{eq:triangle} applied to $x_{j+4}$, $x'_j$ and $x'_{j+2}$, we have $x'_{j} \sim x'_{j+2}$.  Hence, $X_{3,j}^1 \join X_{3,j+2}^1$.  By a similar argument, for all distinct $t,r$ in $\{j, j+2, j+4\}$, we have  $X_{3,t}^1 \join X_{3,r}^1$. 

Let us prove that $X_{3,j}^0 \cojoin X_{3,j+2}$. Otherwise,  consider $x'_j \in X_{3,j}^0$ and $x'_{j+2} \in X_{3,j+2}$ such that $x'_j \sim x'_{j+2}$.  By definition of $X_{3,j}^0$, $x'_j \not\sim x_{j+2}$.  Since $\{x_j, x'_j, x'_{j+2}, x_{j+2}\}$ cannot induce a $C_4$, $x_j \sim x'_{j+2}$. By~\eqref{eq:triangle} applied to $x_{j}$, $x_{j+4}$ and $x'_{j+2}$, we have $x'_{j+2} \sim x_{j+4}$.  By~\eqref{eq:triangle} applied to $x'_{j+2}$, $x_{j+4}$ and $x'_{j}$, we have  $x_{j+4} \sim x'_{j}$.  By~\eqref{eq:triangle} applied to $x_{j+4}$, $x'_{j}$ and $x_{j+2}$, we have  $x'_{j} \sim x_{j+2}$, a contradiction.  Hence, $X_{3,j}^0 \cojoin X_{3,j+2}$.  By a similar argument, for all distinct $t,r$ in $\{j, j+2, j+4\}$, we have  $X_{3,t}^0 \cojoin X_{3,r}$. \qed

\medskip

If the three sets $X_{3,j}$, $X_{3,j+2}$ and $X_{3,j+4}$  satisfy the second conclusion of Observation~\ref{obs:triangular}, we say that they are in a {\it triangle configuration}.  Otherwise, they are in a \emph{sparse configuration}. Note that if at least one of $X_{3, j}$, $X_{3, j+2}$ or $X_{3, j+4}$ is empty, then the sets are in a sparse configuration (recall that empty set is anticomplete to any set of vertices).


%
\section{Proof of Theorem \ref{thm:main-bounded}}\label{sec:main}

Recall that $$
\mathcal P = \left(\{c_1\}, \dots, \{c_6\}, X_6, X_2, X_{4, 1}^0, X_{4, 1}^1, \dots, X_{4, 6}^0, X_{4, 6}^1, X_{3, 1}, \dots, X_{3, 6}\right).
$$
By Observations~\ref{obs:AdjX6}, \ref{obs:X2-uniform}, \ref{obs:X4Mon} and \ref{obs:X3Mon}, $\mathcal P$ is a monotone partition. Let us prove that it has the hev-property. 

Let $Y$ be a set among $\{c_1\}, \dots, \{c_6\}$. 
By the definition of the sets $X_{i, j}$, $Y$ is uniform to all sets of $\mathcal P\setminus Y$.  
Also, by Observations~\ref{obs:AdjX6} and~\ref{obs:X2-uniform}, every set $Y\in \{X_2, X_6\}$ is uniform to all sets of $\mathcal P\setminus Y$.  
Consider now a set $Y$ among $X_{4, 1}^0, X_{4, 1}^1, \dots, X_{4, 6}^0, X_{4, 6}^1$. 
By Observation~\ref{obs:X4Mon}, $Y$ is uniform to all sets of $\mathcal P\setminus Y$, except possibly one.  
Hence, by Lemma~\ref{l:deg1B}, it is enough to prove that the partition  $(X_{3, 1}, \dots, X_{3, 6})$ has the hev-property (for $G[X_3]$). 

Since being $(4K_1, C_4, P_6)$-free is a hereditary property, it is enough to prove that some extreme vertex of some $X_{3, i}$ is uniform to all sets $X_{3, j}$, $j\neq i$ (we mean that there is no need to check all subsets  of $V(G)$).

Suppose that $X_{3,1}$, $X_{3,3}$ and $X_{3,5}$ are not in a sparse configuration. This means that they are all non-empty, and in a triangle configuration. 
If some set $X_{3, i}^0$ ($i \in \{1, 3, 5\}$) is non-empty, then some minimal vertex in such a set is uniform to all sets   $X_{3, j}$, $j\neq i$  (because of Observations~\ref{obs:X3-join-X3} and \ref{obs:triangular}). 
So, we may assume that all sets $X_{3, i}^0$ ($i \in \{1, 3, 5\}$) are empty, meaning that the $X_{3, i}$'s ($i \in \{1, 3, 5\}$) are complete to each other. 
Hence, again, any vertex in such a set  $X_{3, i}$ is maximal and uniform to all sets $X_{3, j}$, $j\neq i$.
Hence, we may assume that $X_{3,1}$, $X_{3,3}$ and $X_{3,5}$ are in a sparse configuration.  
By a symmetric argument, $X_{3,2}$, $X_{3,4}$ and $X_{3,6}$ are also in a sparse configuration. 
Hence, by the the definition of the sparse configuration and Observation~\ref{obs:X3-join-X3}, every set $X_{3, i}$ is uniform to all sets $X_{3, j}$ ($j\neq i$) except possibly one. 
So, the partition has the hev-property by Lemma~\ref{l:deg1B}.

We just proved that $\mathcal P$ has the hev-property. Since $\mathcal P$ contains 26 sets, by Theorem~\ref{th:bcw}, the clique-width of any $(4K_1, C_4, P_6)$-free graph which contains a $C_6$ is at most 27.

\section{Conclusion}\label{sec:conclusions}

In this paper, we consider the complexity of colouring ($4K_1,C_4$)-free graphs.   
Lozin and  Malyshev \cite{Lozin} conjectured that the problem can be solved in polynomial time.  
Even though the problem is still open, we show there is a polynomial-time algorithm for	colouring ($4K_1,C_4,P_6$)-free graphs. 
We note that the following classes of graphs also admit polynomial-time colouring algorithms: ($4K_1,C_4,C_5$)-free graphs as given in \cite{FraHam2018}, ($4K_1,C_4,C_6$)-free graphs as given in \cite{Pen2020}. 

Our method to bound the clique-width generalizes some previous works. 
The {\it box} graph $b(G;\mathcal{P})$ of the graph $G$ with respect to a partition $\mathcal P = (A_1, \dots, A_k)$ is the graph whose vertices are the sets $A_i$, and two vertices $A_i, A_j$ of $b(G;\mathcal{P})$ are adjacent if
and only if $A_i$ is not uniform to  $A_j$. 

The partition ${\cal P}$ is \emph{$k$-near-uniform} if for all $i\in \{1, \dots, k\}$, $A_i$ is a clique\footnote{We note that the definition of a $k$-near-uniform partition in \cite{FraHam2017} is incomplete. The sets $S_i$'s must be cliques for the next theorem to hold.}, and $b(G;\mathcal{P})$ is a graph with maximum degree one (so a matching plus some isolated vertices). The key to prove Theorem~\ref{thm:C7} in~\cite{FraHam2017} is the following: 

\begin{Theorem}[\cite{FraHam2017}]
	\label{thm:near-uniform}
	Let $G$ be a graph admitting a $k$-near-uniform  partition $\mathcal P = (A_1, \dots, A_k)$ such that if $A_i$ is not uniform to $A_j$, then $G[A_i \cup A_j]$ is $C_4$-free. Then, the clique-width of $G$ is at most $2k$. 
\end{Theorem}

We observe that by applying  Lemma~\ref{l:uniMon}, a $k$-near-uniform partition as in the theorem is in fact monotone (because any set of the partition is uniform to all other sets except possibly one, and in this case, we use the fact pointed out in Theorem~\ref{th:monotone}: two cliques in a $C_4$-free graph are monotone to each other). Then, by applying Lemma~\ref{l:deg1B} repeatedly, we can see that such a partition has the hev-property. Hence, Theorem~\ref{th:bcw} shows that the bound $2k$ in Theorem~\ref{thm:near-uniform} can be improved to $k+1$. 

Also, our approach allows some generalizations. For instance, to prove the hev-property, it is enough that $b(G;\mathcal{P})$ is a forest (instead of a matching):

\begin{Theorem}\label{thm:box}
	Let $G$ be a graph and $\mathcal P = (A_1, \dots, A_k)$ be a monotone partition of $V(G)$.  
	If $b(G;\mathcal{P})$ is a forest, then $\mathcal P$ has the hev-property (in particular, $G$ has clique-width at most $k+1$ whenever the sets $A_i$ are all cliques).
\end{Theorem}

\noindent{\it Proof.}
Since $b(G;\mathcal{P})$ is a forest, it can be obtained by repeatedly adding vertices of degree at most~1, which means that the hev-property can be proved by several applications of Lemma~\ref{l:deg1B}. The clique-width is then at most $k+1$ by Theorem~\ref{th:bcw}.
\qed

\medskip

Note that Theorem~\ref{thm:box} is in some sense best possible, since for all integers $k\geq 3$, the first and last authors of the present paper found a construction of graphs of unbounded clique-width with a monotone partition into $k$ cliques, whose corresponding box graph is a chordless cycle of length~$k$, see~\cite{DBLP:journals/dm/HoangT24}. 
We propose the following conjectures to investigate further these matters.  We denote the clique-width of $G$ by $\cwd(G)$.

\begin{Conjecture}
	There exists a function $f$ such that the following holds.
	If $G$ is a graph, ${\cal P} = (A_1, \ldots, A_k$) is
	a monotone partition of $V(G)$ into cliques and for
	every chordless cycle $A_{i_1} \dots A_{i_c} A_{i_1}$ of
	$b(G;\mathcal{P})$, $\cwd(G[A_{i_1} \cup \dots \cup A_{i_c}]) \leq x$,
	then $$\cwd(G) \leq f(k, x).$$
\end{Conjecture}

\begin{Conjecture}
	There exists a function $g$ such that the following holds.
	If $G$ is a graph and ${\cal P}$ is a monotone partition of $V(G)$ into cliques such that $b(G;\mathcal{P})$ is a chordless cycle, then
	$\cwd(G) \leq g(k) \log |V(G)|$.
\end{Conjecture}

\section*{Acknowledgements}

This work was supported by the Canadian Tri-Council Research Support Fund. The author  C.T.H. was supported by individual NSERC Discovery Development Grant no. DDG-2024-00015.  
Most of this work was done during the stay of the author R.J. in Lyon, with generous support of ``Fonds Recherche de l'ENS de Lyon''. 
The author N.T. is supported by Projet ANR GODASse, Projet-ANR-24-CE48-4377.

\end{document}